\documentclass[twoside]{article}
\usepackage{fleqn,espcrc2}

\usepackage{graphicx}
\usepackage{epsfig}
\usepackage[figuresright]{rotating}

\newcommand{\AmS}{{\protect\the\textfont2 A\kern-.1667em\lower.5ex\hbox{M}\kern-.125emS}}
\newcommand{\E}[1]{10$^{#1}$eV}

\title{Note on the Origin of the Highest Energy Cosmic Rays.}

\author{Pierre Billoir, Antoine Letessier-Selvon\address{Laboratoire de Physique Nucl\'eaire et de Haute \'Energie, \\ 
        CNRS-IN2P3, Universit\'es Paris VI-VII,\\
Tour 33 rez de Chaus\'ee,\\
75252 Paris Cedex 05 - France.}}

\begin{document}

\begin{abstract}
In this note we argue that the galactic model chosen by E.-J.~Ahn, G.~Medina-Tanco, P.L.~Bierman and T.~Stanev 
in their paper discussing the origin of the highest energy cosmic rays, is alone responsible for the focusing of positive particles towards 
the North galactic pole. We discuss the validity of this model, in particular in terms of field reversals and radial extensions.
We conclude that with such a model one cannot retreive any directional information from the observed direction of the cosmic rays.
In particular one cannot identify point sources at least up to energies of about 200~EeV\footnotemark\ . Therefore the apparent clustering of the back-traced
highest energy cosmic rays observed to date cannot be interpreted as an evidence for a point source nor for the identification of M87, 
which happens to be close to the North pole, as being such a source.
\vspace{1pc}
\end{abstract}

\maketitle
\footnotetext{EeV for Exa electron Volts or \E{18}}
It was recently claimed \cite{eux} that the Ultra High Energy Cosmic Rays (UHECR)
observed up to now, may originate from a common source located in the direction of the
North pole of our Galaxy; the M87 cluster (Virgo) being 
the best candidate for such an emitting object. This conclusion  was drawn from the 
result of the back-tracing of the 13 highest energy cosmic rays observed to date (assumed to be protons, or helium nuclei for the
most energetic ones) in a modeled galactic magnetic field. In their study the field originates from a
``galactic wind'' (analogous to the solar wind) extending far away from the
visible part of the galactic disk.
\par In this paper we study the mathematical properties of such a 
field, and show that the observed convergence is an intrinsic property of 
the field model, and not an evidence of a unique, pointlike source 
for the highest energy cosmic rays.

\section{Focussing properties of the proposed magnetic wind}
As far as ultra-high energies (over $100$~EeV) are concerned, the bending effects
are dominated by the long range behaviour of the galactic field. In the model
proposed in \cite{eux}, the asymptotic field, in spherical coordinates, is purely azimuthal and reads~: 
$$B_{\varphi}=B_{\odot}r_{\odot}\frac{\sin \theta}{r}$$
where the normalization constant $B_{\odot}r_{\odot}=70~\mu$G.kpc is defined from the local value of the field in
the Solar system. The bounds of the region where
the galactic wind extends are not well defined; the authors use 1.5 Mpc in
their numerical simulations.  
\par The most important feature of such a field (in the absence of a cutoff 
on $r$) is that the bending integral
$\int {B_\varphi~dr}$ is divergent in any radial 
direction except the polar axis. As a result, whatever the energy, a charged particle can never
escape to infinity in a direction other than a pole. In practice, this strong
focussing effect is limited by the cutoff, which therefore plays a crucial role.
Using  the field and radial limits given above the bending integral is about
$500\times \sin \theta$~EeV along a radial trajectory. In other words 
particles of $100$~EeV will only escape within a cone of less than about 10 degrees around the polar axis.

\par When considering particles of a given sign
the orientation of the radial component of the Lorentz force depends on the
polar projection of the velocity, therefore   
the ``positive'' pole (as defined by the orientation of curl $\vec{B}$) is focussing, while the ``negative'' pole
is antifocussing.

\par These features are intrinsic to the field model 
(especially its slow decrease with the distance) 
and we can suspect that the convergence of the 
trajectories found in \cite{eux} is therefore not related to any specific property
of the observed set of highest energy cosmic rays.     

\section{Numerical simulations with random data}  
To confirm this hypothesis, we
have numerically back-traced, in the field model of Ref.~\cite{eux}, 
a {\it \underline{random}} set of cosmic rays drawn from a uniform distibution on the (Earth) sky. 
As most of the observered high energy events are actually concentrated around $10^{20}$ eV, and since the authors of~\cite{eux} 
have further assumed that the two highest energy events could be Helium nuclei the range of magnetic rigidity of the
observed data sample is quite narrow. For a fair comparison we have therefore generated a sample of protons with a flat energy distribution 
ranging from 100 to 160~EeV.
\begin{figure}[t]
\begin{minipage}{\textwidth}
\begin{minipage}[t]{0.5\textwidth}
\vspace*{-1cm}
\epsfig{file=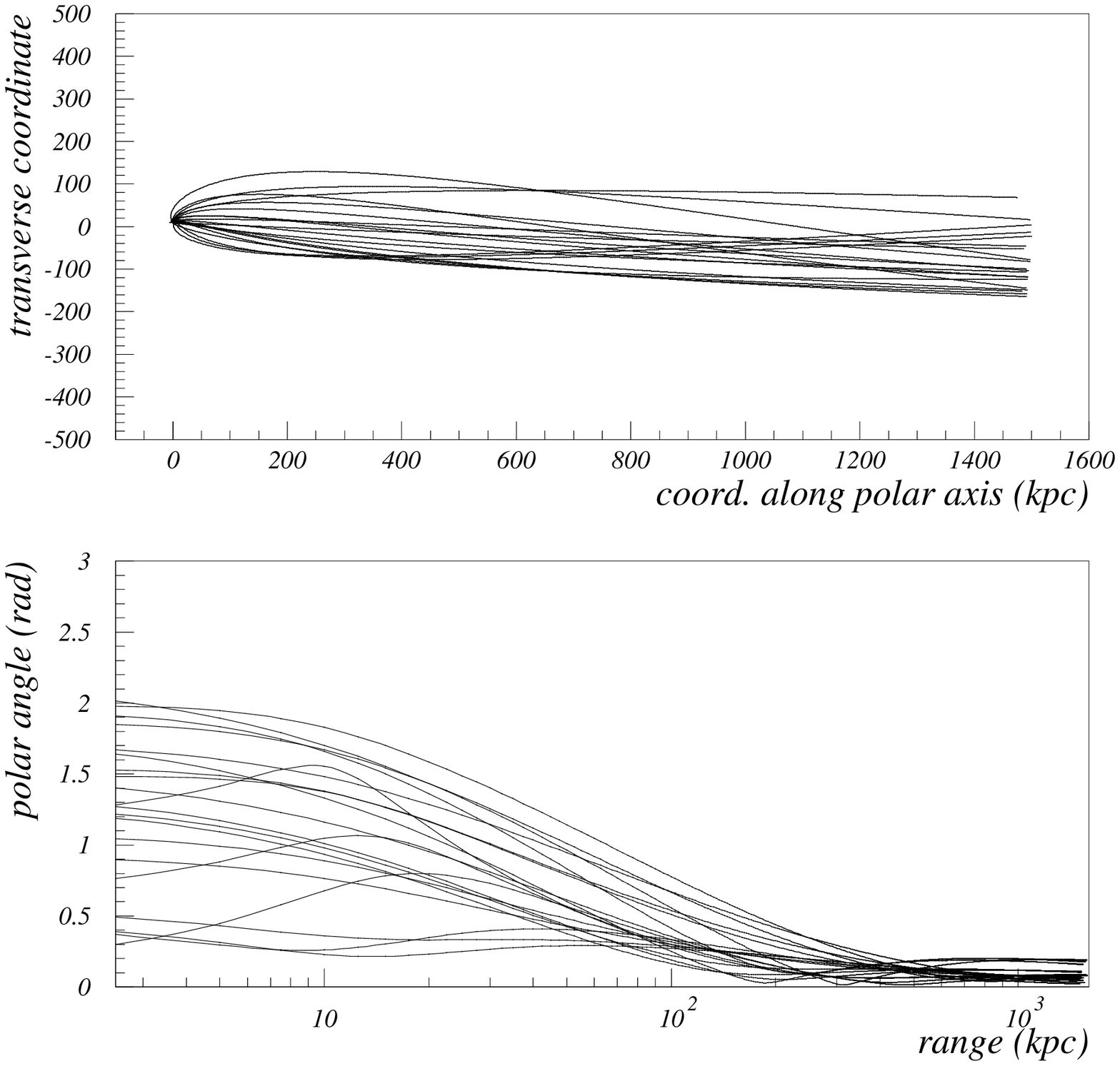,width=80mm} 
\end{minipage}
\begin{minipage}[t]{0.5\textwidth}
\vspace*{-1cm}
\begin{flushright}
\epsfig{file=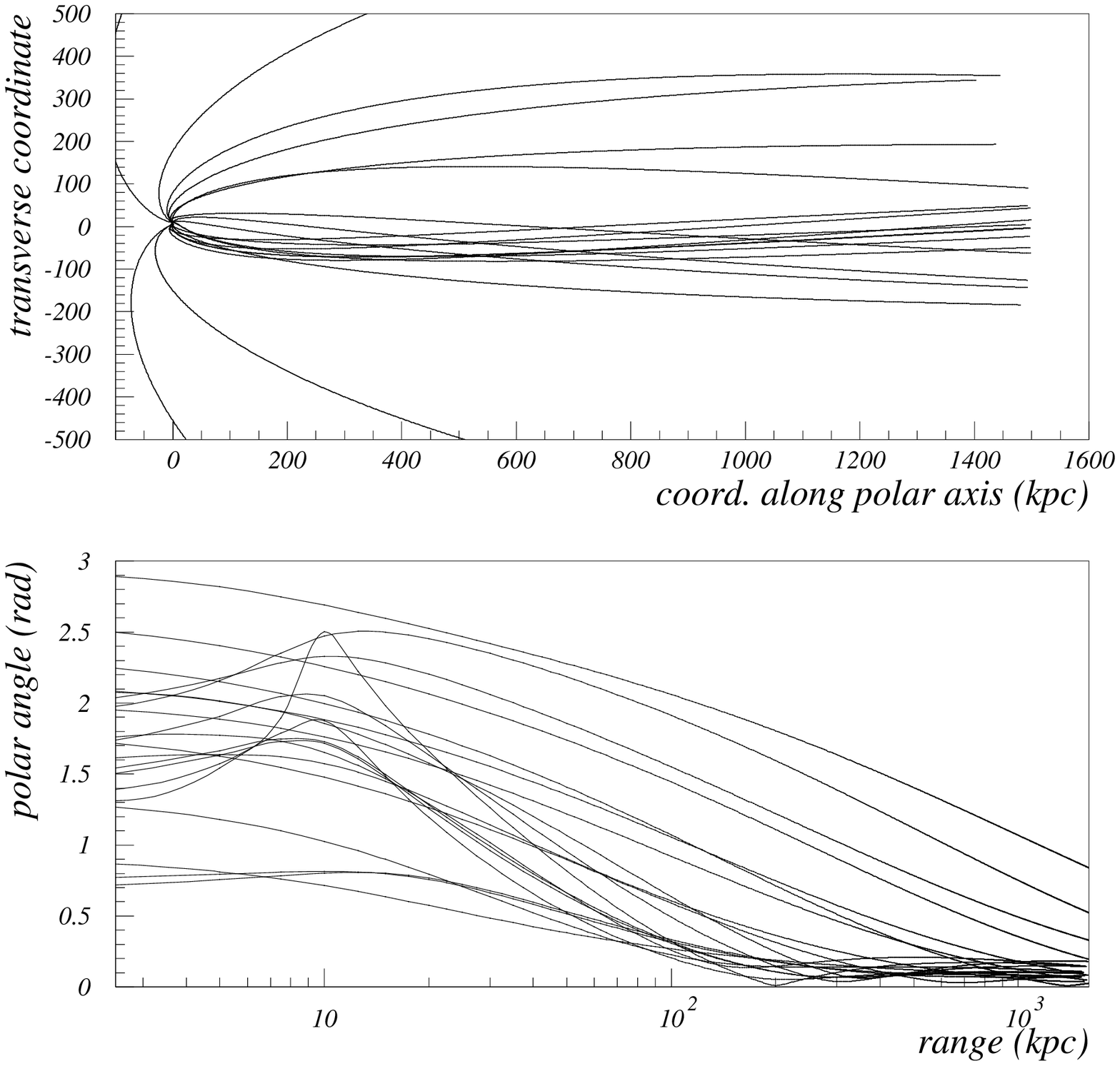,width=80mm} 
\end{flushright}
\end{minipage}
\vspace*{-1.2cm}
\caption{Trajectory of protons of 100 to 160 EeV, back-traced from the Earth's northern (left) and southern (right) hemispheres, 
the Galactic North pole is to the right. Top~:~Transverse coordinate projected on the galactic 
meridian plane containing the Earth versus range. Bottom~:~Polar angle versus range.}
\label{fig:1}
\end{minipage}
\vspace{-0.5cm}
\end{figure}

As expected the trajectories drawn on Fig.\ref{fig:1} clearly  show the strong focussing effect of the field.
Most of the focussing take place over the first few 100 kpc as mentioned in \cite{eux}. We have drawn separately
the trajectory for particle reaching the Earth's northern hemisphere and reaching the southern one. The focussing effect is stronger 
and faster for rays originating from the north\footnote{The 13 events used in \cite{eux} were all observed in the northern hemisphere, which is where
the largest detectors are installed.} which is expected given the configuration of the Earth's rotation axis
with respect to the galactic center. Using a random population we obtain the same behavior as in \cite{eux} showing
that the focussing is a property of the field and not of the events.

\begin{figure}[t]
\vspace*{8.24cm}
\end{figure}

\begin{figure}[htb]
\begin{flushleft}
\epsfig{file=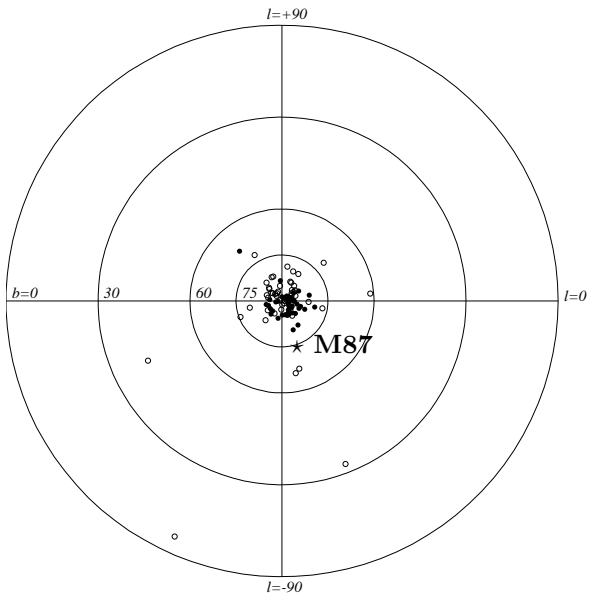,width=80mm,bbllx=50pt,bblly=50pt,bburx=520pt,bbury=520pt,clip=}
\end{flushleft}
\vspace*{-1.3cm} 
\caption{A representation similar to the plot of Ref~\cite{eux}. Black (open) circles are for rays hitting the Earth's northern (southern) hemisphere.}
\label{fig:2}
\vspace*{-10cm}
\setlength{\unitlength}{1mm}  
\begin{picture}(100,100)
\put(39,47.4){\bf $\star$~M87}
\end{picture}
\vspace*{-0.8cm}
\end{figure}

\section{Discussion and conclusion}
First the validity of the model may be questioned: if there are reversals of
the azimuthal component of $B$ in the galactic disk (as acknowledged in 
\cite{eux}), how can the wind remain consistent  with
a ``coherent'' parametrization $\sin \theta / r$ at long distances (several
100 kpc, much more than the distance between regions with reversed fields) ? 
Normally one would expect some destructive interferences between the wind 
contributions coming from differents parts of the disk, hence an intensity 
decreasing faster than $1/r$;
the argument that most cosmic rays are observed in the direction opposite
to the galactic center (where no reversal occurs) is not valid, because their 
bending depends mainly on the long range behaviour of the field.
\par  If however the model of Ref.\cite{eux} is true, then the accumulation of events at the pole
is not at all an evidence for a pointlike
source of the observed rays. This model only demonstrates that our 
sensitivity on extragalactic charged particles might be limited to a small solid angle 
(decreasing with energy) around the galactic polar direction, whatever their initial
origin could be.
\par One should note that even if the extragalactic flux is isotropic the integrated 
luminosity at Earth would be the same with or without this galactic wind. Despite the strong dispersion in the original directions
the restriction of the angular acceptance due to the focussing effect is compensated by the collecting area.

\par Addressing the question of the active galaxy M87 (Virgo A) as a possible source of UHECR, 
the only possible affirmative conclusion is: {\em if} the field
model of \cite{eux} is valid, {\em and if} the sources are known pointlike objects, a possible 
candidate is M87.  However, as acknowledged in~\cite{eux}, this scenario implies
a regular transverse magnetic field from here to Virgo, i.e. over a length of about 20 Mpc, of about 2 nanogauss.
Therefore the overall system would behave like a spectrum analyzer : a magnetic spectrometer 
(the transverse field) followed by a collimator (the Galactic field), strongly selecting the initial momentum 
of the cosmic rays.

\paragraph{Acknowledgements} We thank P.~Astier, X.~Bertou, M.~Boratav and M. Lemoine for their usefull comments and fruitful discussions.


\begin{thebibliography}{9}
\bibitem{eux} Eun-Joo Ahn, Gustavo Medina-Tanco, Peter L. Beirmann, and Todor Stanev,
``The Origin of the highest energy cosmic rays. Do all roads lead back to Virgo?''
astro-ph/9911123 (8 Nov 1999).

\end{thebibliography}
\end{document}